\documentclass[12pt,a4paper]{article}
\pdfoutput=1

\usepackage[utf8]{inputenc}
\usepackage[T1]{fontenc}

\usepackage{cite}
\usepackage{amsmath}
\usepackage{color}
\usepackage{amsfonts}
\usepackage{amssymb}
\usepackage{graphicx}
\usepackage{geometry}
\usepackage{amssymb,epsfig,subfigure}
\usepackage{hyperref}
\usepackage{comment}
\usepackage[font=footnotesize]{caption}

\makeatletter
\renewcommand\section{\@startsection {section}{1}{\z@}%
                                 {-3.5ex \@plus -1ex \@minus -.2ex}
                                   {2.3ex \@plus.2ex}%
                                   {\normalfont\large\bfseries}}
\renewcommand\subsection{\@startsection{subsection}{2}{\z@}%
                                   {-3.25ex\@plus -1ex \@minus -.2ex}%
                                     {1.5ex \@plus .2ex}%
                                     {\normalfont\bfseries}}
\renewcommand\subsubsection{\@startsection{subsubsection}{3}{\z@}%
                                   {-3.25ex\@plus -1ex \@minus -.2ex}%
                                     {1.5ex \@plus .2ex}%
                                     {\normalfont\itshape}}
\makeatother

\def\pplogo{\vbox{\kern-\headheight\kern -29pt
\halign{##&##\hfil\cr&{\ppnumber}\cr\rule{0pt}{2.5ex}&\ppdate\cr}}}
\makeatletter
\def\ps@firstpage{\ps@empty \def\@oddhead{\hss\pplogo}%
  \let\@evenhead\@oddhead 
}
\def\maketitle{\par
 \begingroup
 \def\thefootnote{\fnsymbol{footnote}}
 \def\@makefnmark{\hbox{$^{\@thefnmark}$\hss}}
 \if@twocolumn
 \twocolumn[\@maketitle]
 \else \newpage
 \global\@topnum\z@ \@maketitle \fi\thispagestyle{firstpage}\@thanks
 \endgroup
 \setcounter{footnote}{0}
 \let\maketitle\relax
 \let\@maketitle\relax
 \gdef\@thanks{}\gdef\@author{}\gdef\@title{}\let\thanks\relax}
\makeatother

\numberwithin{equation}{section}

\newcommand\eea{\end{eqnarray}}
\newcommand\bea{\begin{eqnarray}}

\def\beq{\begin{equation}}
\def\eeq{\end{equation}}

\def\e{{\rm e}}

\newcommand{\be}{\begin{equation}}
\newcommand{\ee}{\end{equation}}
\newcommand{\ba}{\begin{align}}
\newcommand{\ea}{\end{align}}
\newcommand{\bg}{\begin{gather}}
\newcommand{\eg}{\end{gather}}
\newcommand{\bseq}{\begin{subequations}}
\newcommand{\eseq}{\end{subequations}}

\renewcommand{\tanh}{\mathop{\rm th}\nolimits}

\newcommand{\Tr}{{\rm Tr}}

\renewcommand{\t}{\tilde}

\newcommand{\tr}{{\rm tr}}
\newcommand{\mc}{\mathcal}

\begin{document}
\setcounter{page}0
\def\ppnumber{\vbox{\baselineskip14pt
}}
\def\ppdate{
} \date{}

\author{Horacio Casini, Raimel Medina, Ignacio Salazar, Gonzalo Torroba\\
[7mm] \\
{\normalsize \it Centro At\'omico Bariloche and CONICET}\\
{\normalsize \it S.C. de Bariloche, R\'io Negro, R8402AGP, Argentina}
}

\bigskip
\title{\bf  Renyi relative entropies and \\ renormalization group flows
\vskip 0.5cm}
\maketitle

\begin{abstract}
Quantum Renyi relative entropies provide a one-parameter family of distances between density matrices, which generalizes the relative entropy and the fidelity. We study these measures for renormalization group flows in quantum field theory. We derive explicit expressions in free field theory based on the real time approach. Using monotonicity properties, we obtain new inequalities that need to be satisfied by consistent renormalization  group trajectories in field theory. These inequalities play the role of a second law of thermodynamics, in the context of renormalization group flows.
Finally, we apply these results to a tractable Kondo model, where we evaluate the Renyi relative entropies explicitly. An outcome of this is that Anderson's orthogonality catastrophe can be avoided by working on a Cauchy surface that approaches the light-cone.
\end{abstract}
\bigskip

\newpage

\tableofcontents

\vskip 1cm

\section{Introduction}\label{sec:intro}

Quantum field theory (QFT) describes the long-distance limit of many systems of interest in high energy and condensed matter physics. Novel collective phenomena are often observed at strong coupling, and a long-term goal is to develop tools to understand strongly interacting QFTs. In this direction, we have witnessed important recent progress by applications of results from quantum information theory (QIT). By studying how degrees of freedom are entangled, and how this changes from microscopic to macroscopic scales, new results on the nonperturbative behavior of QFTs have been obtained. These range from ansatze for ground state wavefunctions, to irreversibility of the renormalization group and insights into quantum gravity.

\begin{figure}[h!]
\begin{center}  
\includegraphics[width=0.7\textwidth]{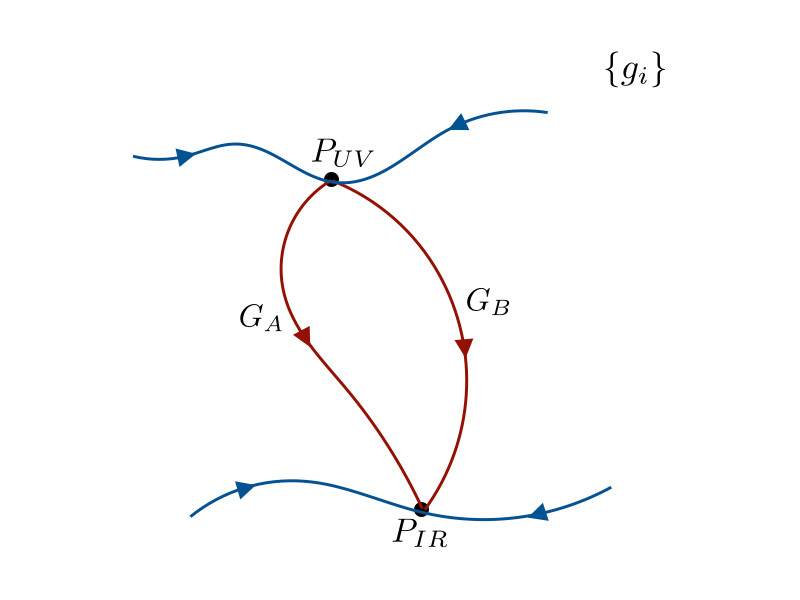}
\captionsetup{width=0.9\textwidth}
\caption{Possible RG trajectories between two fixed points, following~\cite{Wilson:1973jj}. Red lines denote relevant trajectories from $P_{UV}$, while blue lines are associated to irrelevant deformations.
}
\label{fig:RG}
\end{center}  
\end{figure}

In this work we will focus on certain nonperturbative aspects of the renormalization group (RG). The RG gives flows or trajectories in the space of couplings $\lbrace g_i \rbrace $ as a function of some distance or energy scale. These flows generically include fixed points, as well as relevant and irrelevant trajectories~\cite{Wilson:1973jj}. This is illustrated in Fig.~\ref{fig:RG}, which shows two fixed points $P_{UV}$ and $P_{IR}$, relevant flows from $P_{UV}$ to $P_{IR}$ (in red), and irrelevant flows (in blue).

By now, it has been established that the RG for unitary relativistic QFTs is irreversible in two~\cite{Zamolodchikov:1986gt, Casini:2004bw}, three~\cite{Casini:2012ei} and four space-time dimensions~\cite{Komargodski:2011vj, Casini:2017vbe}. This means that one can associate an intrinsic quantity $C$ to a fixed point, and a necessary condition to connect two fixed points by the RG is that
\be\label{eq:Cirrev}
C_{UV} > C_{IR}\,.
\ee
This can be understood in terms of the entanglement entropy (EE),
\be
S(\rho_V) = - \Tr\,\rho_V \,\log \rho_V
\ee
where
\be\label{eq:vacrho}
\rho_V= \Tr_{\bar V}\,|0\rangle \langle 0|
\ee
is the vacuum reduced density matrix obtained by tracing over the degrees of freedom in the complement of a region $V$. The irreversibility of the RG is a consequence of unitarity and strong subadditivity of the entanglement entropy~\cite{Casini:2004bw, Casini:2012ei, Casini:2017vbe}.\footnote{So far, it is not known whether this holds for space-time dimensions $d \ge 5$.}

Here we are interested in the following question: given two fixed points that satisfy (\ref{eq:Cirrev}), which RG flows between them are possible? In particular, we would like to establish necessary conditions that must be satisfied by consistent RG trajectories in unitary QFTs. For this, it is natural to analyze the RG in terms of a quantum information distance between two reduced density matrices: $\sigma_V$, characterizing the UV fixed point, and $\rho_V$, associated to the theory that undergoes the flow. One very useful notion of distance is the relative entropy,
\be\label{eq:rel}
S(\rho||\sigma)=\Tr\,\rho(\log \rho - \log \sigma)\,,
\ee
which measures the distinguishability between the two states. The relative entropy is positive and monotonic under increasing the size of the region. Based on this, Ref.~\cite{Casini:2016fgb} proved the entropic version of irreversibility of boundary RG flows, and~\cite{Casini:2016udt} gave an alternative proof of the $c$-theorem. Another notion of distance is the quantum fidelity,
\be\label{eq:Fid}
F(\rho, \sigma) =\Tr\,\sqrt{\sigma^{1/2} \rho \sigma^{1/2}}\,,
\ee
which reduces to the overlap between wavefunctions when the states $\rho$ and $\sigma$ are pure. Such overlaps appear for instance in the evaluation of the boundary entropy in terms of boundary states~\cite{Cardy:2004hm}. 

In this work we will analyze a family of distance measures called quantum Renyi relative entropies,
\be\label{eq:Renyi}
S_\alpha(\rho||\sigma)=-\frac{1}{1-\alpha}\,\log\,\Tr\,\left(\sigma^{\frac{1-\alpha}{2\alpha}}\rho \sigma^{\frac{1-\alpha}{2\alpha}} \right)^\alpha\,.
\ee
Although other definitions of measures exist in the QIT literature (e.g.~\cite{petz1986quasi}), our main motivation for studying (\ref{eq:Renyi}) is that the $S_\alpha$ interpolate between (\ref{eq:rel}) and (\ref{eq:Fid}) as $\alpha$ varies between $1$ and $1/2$. This fact, together with monotonicity properties satisfied by the $S_\alpha$, will allow us to derive new necessary conditions for consistent RG flows.

The nonlinear dependence of $S_\alpha(\rho||\sigma)$ on $\rho$ and $\sigma$ makes explicit calculations quite hard, and there has not been much work on understanding the physical content of these measures. Some previous works include~\cite{Lashkari:2014yva}, where the $S_\alpha$ were evaluated in 2d CFTs using the replica trick, and~\cite{Bernamonti:2018vmw}, which studied the quantum Renyi divergences~\cite{petz1986quasi} for excited CFT states. Furthermore, the paper~\cite{May:2018tir} studies quantum Renyi divergences to second order in perturbation theory using holography. Some generalizations of Renyi relative entropies to von Neumann algebras have been analyzed in~\cite{Berta:2018ecp, jencova2016r, jenvcova2017renyi}.

In this paper we compute $S_\alpha$ in free field theories using real time methods (Sec.~\ref{sec:real}). We next consider in Sec.~\ref{sec:RG} general consequences of the monotonicity properties of $S_\alpha$ for RG flows. By focusing on the light-cone limit studied in~\cite{Casini:2016fgb, Casini:2016udt}, we show that the $S_\alpha$, which characterize the full RG trajectory, are bounded by quantities intrinsic to the fixed points, such as the boundary entropy or the central charge. Our main results on this, the inequalities~(\ref{eq:bRGbound1}) and (\ref{eq:RGbound2}), closely resemble the second law of thermodynamics $\Delta S \ge \int \frac{dQ}{T}$, where a change in the entropy (a function of state) bounds a quantity that depends on the process. Finally, we illustrate these results in Sec.~\ref{sec:Kondo} with an explicit evaluation of $S_\alpha$ in a tractable Kondo model. In particular, we find that Anderson's orthogonality catastrophe~\cite{anderson1967infrared} can be avoided by computing overlaps of ground states in the light-cone limit; this could be of interest for more general impurity problems in condensed matter physics.

\section{Quantum Renyi relative entropies}
\label{sec:QRRE}

Let us begin by reviewing some basic properties of the $S_{\alpha}$. The quantum Renyi relative entropies (QRRE in what follows) are defined as~\cite{muller2013quantum,wilde2014strong}
\be
S_\alpha(\rho||\sigma)=-\frac{1}{1-\alpha}\,\log\,\Tr\,\left(\sigma^{\frac{1-\alpha}{2\alpha}}\rho \sigma^{\frac{1-\alpha}{2\alpha}} \right)^\alpha\,.
\label{renyi}
\ee
They appear as a natural generalization of the Renyi relative entropies \cite{renyi1961measures} that includes the quantum non-commutativity of the density matrices involved.

In this work we focus mostly on the range $\frac{1}{2}\le\alpha \le 1$. In particular the edges of this interval are characterized by previously known quantum information measures. When $\alpha=1/2$ we have the fidelity distance,
\bea
S_{1/2}(\rho||\sigma)&=&-2 \,\log\,\Tr\,\sqrt{\sigma^{1/2}\rho \sigma^{1/2}}=-2\,\log F(\rho,\sigma) \,,
\label{fid1}
\eea
where  $F(\rho,\sigma)$ denotes the quantum fidelity (\ref{eq:Fid}).
Another interesting case is the limit $\alpha\rightarrow1$
\bea
S_{1}(\rho||\sigma)&=&\Tr (\rho \log \rho-\rho \log \sigma)= S(\rho||\sigma)\,.
\eea
Here $ S(\rho||\sigma)$ is the quantum relative entropy (\ref{eq:rel}).
Then, quantum Renyi relative entropies appear, when $\frac{1}{2}\le\alpha \le 1$, as an interpolation between quantum fidelity and quantum relative entropy. As discussed in Sec.~\ref{sec:intro}, this is one of our main motivation for considering (\ref{renyi}), as opposed to other alternative forms such as~\cite{petz1986quasi}

For the fidelity, a useful representation is given by Uhlmann's theorem~\cite{uhlmann1976transition},
which states that
\bea
F(\rho,\sigma)=\max_{\left|\psi \right>,\left|\phi \right>}\left|\left<\psi | \phi\right>\right|\,,
\label{fid2}
\eea
over purifications $|\psi \rangle,\,|\phi\rangle$ of $\rho, \sigma$. Given this result,
properties of the fidelity can be easily proved. For instance,  (\ref{fid2}) makes it clear that the fidelity is symmetric in its inputs $F(\rho,\sigma)=F(\sigma,\rho)$. One can also see that it is bounded $0<F(\rho,\sigma)<1$. If $\rho=\sigma$,  $F(\rho,\sigma)=1$, while  $F(\rho,\sigma)=0$ if and only if $\rho$ and $\sigma$ have support on orthogonal subspaces.

The Renyi relative entropies $S_\alpha$ also admit representations in terms of extremizing quantities. For instance, in~\cite{doi:10.1063/1.4838835} it was shown that
\be\label{eq:min-rep}
\Tr \,\left(\sigma^{\frac{1-\alpha}{2\alpha}}\, \rho\,\sigma^{\frac{1-\alpha}{2\alpha}} \right)^\alpha= \text{min}_{H \ge 0}\,\left(\alpha\,\Tr(H \rho)-(\alpha-1)\,\Tr\,\left(H^{1/2} \sigma^{\frac{\alpha-1}{\alpha}} H^{1/2}\right)^{\alpha/(\alpha-1)} \right)
\ee
for $0<\alpha<1$; the minimum should be replaced by the maximum for $\alpha>1$. A similar representation is derived in~\cite{muller2013quantum}.
These representations are at the basis of the monotonicity properties that we will now review. 

The $S_\alpha$ are monotonically increasing in $\alpha$~\cite{muller2013quantum,beigi2013sandwiched, doi:10.1063/1.4838835}
\be\label{eq:alphamon}
\frac{d}{d\alpha}S_\alpha(\rho||\sigma)\geq0\,.
\ee
Both the fidelity distance and the relative entropy are positive, and equal to zero only when $\rho= \sigma$. Eq.~(\ref{eq:alphamon}) then gives the same properties for the $S_\alpha$,
\be\label{eq:positivity}
S_\alpha(\rho||\sigma) \ge 0\;\;,\;\; S_\alpha(\rho||\sigma) = 0 \;\;\text{for}\;\;\rho= \sigma\,.
\ee
Another important  property is monotonicity when increasing the size of the algebra. If we consider two regions $V \subset \tilde V$, then
\be\label{eq:monoton}
S_\alpha(\rho_V ||\sigma_V) \le S_\alpha(\rho_{\tilde V} ||\sigma_{\tilde V}) \,.
\ee
This result uses (\ref{eq:min-rep}); see e.g.~\cite{doi:10.1063/1.4838835}. This property is intuitive in QFT: the information-theoretic distance $S_\alpha(\rho||\sigma)$ decreases for smaller regions, because there are less operators localized in the region that can be used to distinguish the states.

In the following sections we will study the consequences of these equations for the RG.

\section{Renyi relative entropies in free field theory} \label{sec:real}

In this section we will calculate the QRREs in free QFT. These are the simplest possible models in field theory, and hence provide a natural place to start understanding the $S_\alpha$. Notwithstanding their simplicity, free models provide an interesting setup for QIT measures, where properties of more general QFTs may be recognized.\footnote{For instance, the role of the area law or the connection with anomalies, were recognized early on in calculations of entanglement entropy~\cite{Srednicki:1993im, Holzhey:1994we, Larsen:1994yt}.} Gaussian states also play a prominent role in quantum information theory, quantum optics and atomic physics --see e.g.~\cite{RevModPhys.84.621, serafini2017quantum} for reviews. Some related works on fermionic and bosonic gaussian states and information-theoretic measures include~\cite{Paraoanu:1999zu, wang2000bures,cozzini2006,marian2012uhlmann, banchi2014quantum, banchi2015quantum, Seshadreesan:2017akv}. Furthermore, 

We will work in real time, relating the Gaussian correlators on a fixed Cauchy slice to the density matrix; this procedure is reviewed in~\cite{Casini:2009sr}. This approach is also useful for lattice calculations, and the results will be applied to a Kondo model in Sec.~\ref{sec:Kondo} below. Our results are valid for Gaussian states, which have broader applicability than free QFTs. This approach is presented in Appendix \ref{app:gaussian}. In this section, however, we frame the discussion in terms of free theories.

\subsection{Renyi relative entropies for free fermions}\label{subsec:fermions}

Consider two field theories of fermions, with the same field content, but with different Hamiltonians. We will restrict to free theories (quadratic Hamiltonians), which lead to Gaussian ground states. In the present derivation we work at zero temperature and vanishing chemical potential, but do not require Poincar\'e invariance. Let us denote the two different Hamiltonians on the lattice by
\be
H = \sum_{ij} \, M_{ij}\,\psi^\dag_i \psi_j\;\;,\;\;H'= \sum_{ij} \, M_{ij}'\,\psi^\dag_i \psi_j\,.
\ee
In the standard case, these Hamiltonians arise from discretizing the theories of interest on a constant time Cauchy surface. But let us point out from the start that we will also be interested in more general Cauchy surfaces. In particular, in relativistic theories below, the appropriate Cauchy surfaces will approach the light-cone limit.

The reduced density matrices are denoted by $\sigma_V$ and $\rho_V$, respectively; as in (\ref{eq:vacrho}), these are obtained by tracing over the fermions on the sites in the complement of the set $V$. Our goal is to compute $S_\alpha(\rho||\sigma)$.\footnote{When it does not lead to confusions, we will avoid the subscript `$V$'.}

The fermion modes obey $\lbrace \psi_i, \psi_j^\dag \rbrace= \delta_{ij}$.
The non-vanishing two-point correlators on the Cauchy surface are given by the zero-temperature Fermi-Dirac distribution,
\be
\langle \psi_i \psi^\dag_j \rangle = C_{ij}\;,\;\langle \psi_i^\dag \psi_j \rangle = \delta_{ij}- C_{ji}
\ee
with $C = \Theta(-M)$. Similar expressions hold for the other theory, with $C' = \Theta(-M')$.

Consistently with Wick's theorem, the reduced density matrix is given by a Gaussian state~\cite{peschel2003, Casini:2009sr}
\be
\rho_V = K e^{- \mc H_V}\;,\; \mc H_V = \sum_{ij \in V} H_{Vij} \psi_i^\dag \psi_j
\ee
where $\mathcal{H}_V$, which is known as the modular Hamiltonian, is fixed in terms of the correlator by requiring $\tr (\rho \psi_i^\dag \psi_j) = C_{ij}$. The result is
\be
H_V = - \log( C^{-1}-1 )\,.
\ee
The normalization constant $K = 1/\det(1+ e^{-  H_V})$.

The QRREs can be calculated explicitly because of two key properties. First, for a Gaussian state $\rho$, the power $\rho^\alpha$ is again a Gaussian state, with modular Hamiltonian $\alpha H_{Vij}$. Second, because of the algebra of creation and annihilation operators, the product of two different Gaussian states is again a Gaussian state, whose modular Hamiltonian matrix can be obtained in terms of the Baker-Campbell-Hausdorff (BCH) formula.

In order to see this, it is convenient to introduce Majorana fermions $w_I= (\psi_j + \psi^\dag_j, i (\psi_j - \psi^\dag_j))$, and rewrite the reduced density matrix as
\be
\rho \propto \exp \left( -\frac{i}{4} \sum_{IJ \in V} G_{IJ} w_I w_J \right)\,,
\ee
where $G$ is real and antisymmetric. Then using $ \lbrace w_I,w_J \rbrace = 2 \delta_{IJ}$ obtains~\cite{Brezin}
\begin{equation}\label{eq:prodf}
e^{\frac{i}{4} w^T R_1 w } e^{\frac{i}{4} w^T R_2 w} = \e^{\frac{i}{4} w^T R w }, \hspace{1cm} e^{R_1} e^{R_2} = e^{R}.
\end{equation}
This allows to compute products of Gaussian density matrices, which is what we need to evaluate the $S_\alpha$.
This method was used in~\cite{banchi2014quantum} to compute the fidelity, and a similar approach is presented in the Appendix for the calculation of $S_\alpha$. The final result is
\bea
\label{eq:SfinalF}
S_\alpha(\rho||\sigma) & =& -\Tr \log(1-C) - \frac{\alpha}{1-\alpha}\Tr \log(1-C') \\
& -&\frac{1}{1-\alpha}\Tr \log \left[ 1 +\left( \Big(\frac{C}{1-C} \Big)^{\frac{1-\alpha}{2 \alpha}} \frac{C'}{1-C} \Big(\frac{C}{1-C} \Big)^{\frac{1-\alpha}{2 \alpha}}\right)^{\alpha} \right] \nonumber\,.
\eea
Recall that $C$ is the correlator associated to $\sigma$, while $C'$ is the one associated to $\rho$.

\subsection{Resolvent method}

Eq.~(\ref{eq:SfinalF}) gives a closed expression for $S_\alpha$ in terms of the fermion correlators $C$ y $C'$. 
In order to compute the previous nontrivial powers of operators, it is often convenient to use their resolvents. 

Let us introduce the resolvent of an operator $M$,
\be
R(M, z) = \Tr \left(\frac{1}{M-z}+\frac{1}{z} \right)\,.
\ee
We have added the $1/z$ term compared to the standard definition in order to achieve convergence at large $z$. We can do this because, in all our expressions below, this term will be multiplied by functions that vanish at $z=0$.

The correlators $C$ and $C'$ have eigenvalues between 0 and 1. For a given eigenvalue $\lambda$,
\be
\int_1^\infty\,d\beta \left(\frac{1}{\lambda-\beta}+\frac{1}{\beta} \right)= \log(1-\lambda)\,,
\ee
and thus
\be\label{eq:resC}
\Tr\,\log(1-C) = \int_1^\infty d\beta\,R(C, \beta)\,.
\ee

Next, we focus on the more complicated matrix
\be
M \equiv \left(\frac{C}{1-C}\right)^{\frac{1-\alpha}{2\alpha}}\,\frac{C'}{1-C'}\,\left(\frac{C}{1-C}\right)^{\frac{1-\alpha}{2\alpha}}\,,
\ee
and we need to compute $\Tr\,\log(1+M^\alpha)$.
The matrix $M$ has positive eigenvalues.  For a single eigenvalue $\lambda$, we have
\be
\log(1+\lambda^\alpha)=\frac{1}{2\pi i} \int_{\mathcal C}\left(\frac{1}{z-\lambda}-\frac{1}{z} \right)\,\log(1+z^\alpha)
\ee
where $\mathcal C$ is a contour that runs anti-clockwise around $\lambda$ (it does not contain $z=0$). The term proportional to $1/z$ has vanishing integral, but is added in order to have an integrable integrand at large $z$.
Let us choose the branch cut of $\log(1+z^\alpha)$ to be at $z>0$. We can then deform the contour $\mathcal C$ to run between $(-\infty, 0)$, with the result
\be
\log(1+\lambda^\alpha)=-\frac{1}{2\pi i} \int_0^\infty d\beta\,\left(\frac{1}{\lambda+\beta}-\frac{1}{\beta} \right) \left[\log(1+\beta^\alpha e^{i \pi \alpha}) -\log(1+\beta^\alpha e^{-i \pi \alpha}) \right]\,.
\ee
Therefore,
\be\label{eq:resM}
\Tr\,\log(1+M^\alpha)=-\frac{1}{2\pi i} \int_0^\infty d\beta\,R(M,-\beta)\left[\log(1+\beta^\alpha e^{i \pi \alpha}) -\log(1+\beta^\alpha e^{-i \pi \alpha}) \right]\,.
\ee

Using (\ref{eq:resC}) and (\ref{eq:resM}), the Renyi relative entropies (\ref{eq:SfinalF}) become
\bea
\label{eq:SfinalF2}
S_\alpha(\rho||\sigma) & =& - \int_1^\infty d\beta\,\left(R(C, \beta) + \frac{\alpha}{1-\alpha} R(C', \beta)\right) \\
&+&\frac{1}{1-\alpha} \int_0^\infty \frac{d\beta}{2\pi i}\,R(M,-\beta) \left[\log(1+\beta^\alpha e^{i \pi \alpha}) -\log(1+\beta^\alpha e^{-i \pi \alpha}) \right] \nonumber\,.
\eea
In Sec.~\ref{sec:Kondo} we will apply these results to the case of a Dirac fermion coupled to a Kondo impurity, which undergoes an RG flow.

\subsection{Free bosons}\label{subsec:bosons}

Let us now focus on free bosons. The lattice Hamiltonian is of the form
\be\label{hamilbosons}
H = \frac{1}{2} \sum_i \pi^2_i + \frac{1}{2} \sum_{i,j} K_{ij} \phi_i \phi_j\,,
\ee
where $\phi_i$ and $\pi_j$ obey the canonical commutation relations $\left[ \phi_i, \pi_j \right] = i \delta_{ij}$. We will consider two different Hamiltonians, with quadratic kernels $K$ and $K'$, and evaluate $S_\alpha(\rho||\sigma)$ for their corresponding reduced density matrices. Renyi relative entropies for bosonic gaussian states were also evaluated in~\cite{Seshadreesan:2017akv}.

The two-point functions on the Cauchy surface are parametrized as
\begin{equation}\label{2pointcorr}
\begin{split}
& \langle \phi_i \phi_j \rangle = X_{ij}, \hspace{1cm} \langle \pi_i \pi_j \rangle = P_{ij}  \\
& \langle \phi_i \pi_j \rangle = \langle \pi_j \phi_i \rangle^* = \frac{i}{2}\delta_{ij}.
\end{split}
\end{equation}
with 
\be
X_{ij} = \frac{1}{2} (K^{-\frac{1}{2}})_{ij}, \hspace{1cm} P_{ij} = \frac{1}{2}(K^{1/2})_{ij}\,.
\ee
See e.g.~\cite{Casini:2009sr} for a review of these points. 
The equations in \eqref{2pointcorr} imply the matrices $X$ and $P$ are real Hermitian and positive. Furthermore, introducing
\be\label{eq:Cdef}
C= \sqrt{XP}\,,
\ee
the eigenvalues of $C^2$ are greater than $1/4$.

Consistently with Wick's theorem, the reduced density matrix is given by a Gaussian state of the general form
\be\label{eq:rhoboson-general}
\rho_V \propto \exp\left(-\sum_{i,j \in V} \big( M_{ij} \phi_i \phi_j + N_{ij}\pi_i \pi_j \big) \right)\,.
\ee
This density matrix can be diagonalized by a Bogoliubov transformation, which allows to relate $M$ and $N$ to the correlators (\ref{2pointcorr}). Explicit expressions may be found in~\cite{Casini:2009sr}.

In order to compute the QRREs, 
it is convenient to introduce the variables $Q_I = (\phi_j, \pi_j)$, which satisfy $[Q_I, Q_J] = i \Omega_{IJ}$, where $\Omega = i \sigma_2 \otimes \mathbf 1$ is the symplectic matrix. In this parametrization, the reduced density matrix is of the form
\be
\rho_V \propto \exp \big ( -\frac{1}{2} Q^T G Q \big ),
\ee
where $G$ is real and symmetric, and its blocks are determined by $M$ and $N$ in (\ref{eq:rhoboson-general}). Products of Gaussian states are then given by~\cite{Brezin}
\be\label{eq:prodgauss}
e^{-\frac{1}{2}Q^T G_1 Q}e^{-\frac{1}{2}Q^T G_2Q} = e^{-\frac{1}{2}Q^T G Q}, \hspace{1cm} e^{-i \Omega G_1}e^{-i \Omega G_2} = e^{-i\Omega G}\,.
\ee
This method was used in~\cite{banchi2015quantum} to compute the fidelity for the bosonic case.

Using these properties, in the Appendix we evaluate $S_\alpha$, obtaining
\be\label{Sbosonic}
\begin{split}
S_\alpha (\rho || \sigma) & =\frac{1}{2} \Tr \log \left( \frac{1}{4}-C^2 \right) + \frac{1}{2}\frac{\alpha}{1-\alpha} \Tr \log \left( \frac{1}{4}-C'^2 \right) \\
& + \frac{1/2}{1-\alpha} \Tr \log \left( \left( T^{ \frac{1-\alpha}{2 \alpha} }T'T^{ \frac{1-\alpha}{2 \alpha} } \right)^\alpha - 1 \right),
\end{split}
\ee
where $T$ is given by the following expression
\be
\begin{split}
T = \left( \begin{matrix}
\frac{C^2+\frac{1}{4}}{C^2-\frac{1}{4}} & i \frac{C^2}{C^2-\frac{1}{4}}P^{-1} \\
-i P \frac{1}{C^2-\frac{1}{4}} & P \frac{C^2+\frac{1}{4}}{C^2-\frac{1}{4}}P^{-1}
\end{matrix}
\right).
\end{split}
\ee
(Recall that $C$ is the correlator associated to $\sigma$, while $C'$ is the one associated to $\rho$.) The remaining nontrivial powers of $T$ and $T'$ can be computed explicitly, but the final expressions in terms of the original correlators $(X, P, X', P')$ are rather complicated and will not be presented here. These complications are due to the fact that, unlike the fermionic case, here the $T$ matrices are not block-diagonal, and depend on $P$ (or $X$) as well as on $C$.

\section{Bounds on renormalization group flows} \label{sec:RG}

In this section we analyze some general implications of the monotonicity properties of the $S_\alpha(\rho||\sigma)$ for RG flows. 

For reduced density matrices on a region of typical size $R$, and for a relevant flow with energy scale $m$, the Renyi relative entropies will depend on the dimensionless combination $mR$. The $S_\alpha$ then define a distance that characterizes the RG. The limit $m R \ll 1$ corresponds to the UV, where the relevant deformation flows to zero and $\rho \to \sigma$; in this case, $S_\alpha \to 0$. From (\ref{eq:monoton}), $S_\alpha$ increases with $mR$, signaling an increased distinguishability between the two states. The limit
$m R \gg 1$ parametrizes the IR, where $\rho$ approaches the density matrix of another fixed point. The RG flow will generically be nonperturbative in nature, and $S_\alpha$ will be sensitive to the full trajectory.
The other property that will play an important role is (\ref{eq:alphamon}), which implies that the QRREs with $\alpha<1$ are bounded above by the relative entropy,
\be\label{eq:RGbound1}
S_\alpha(\rho||\sigma) \le S(\rho||\sigma)\;\;,\;\;\frac{1}{2} \le \alpha \le 1\,.
\ee

\subsection{Boundary RG flows}\label{subsec:bRG}

Boundary RG flows occur when a 2D boundary CFT is perturbed by a relevant operator at the boundary $x^1=0$,\footnote{A boundary CFT (BCFT) is defined as a CFT on $x^1>0$, with boundary at $x^1=0$ that preserves half of the conformal symmetries.}
\be\label{eq:bRG}
S= S_{BCFT_{UV}}+ \int dx^0\, g\,\mathcal O\,.
\ee
This triggers a nontrivial RG flow, which we assume ends at a different infrared boundary CFT, $BCFT_{IR}$. A boundary CFT is characterized by an intrinsic quantity known as the boundary entropy $\log g$. It can be obtained as the part of the thermal entropy that is independent of the size of the system~\cite{Affleck:1991tk},
\be
S= \frac{c\pi}{3} \frac{L}{\beta}+ \log g\,.
\ee
This quantity decreases along boundary RG flows~\cite{Affleck:1991tk, Friedan:2003yc}, $\log g_{UV} > \log g_{IR}$, a statement known as the $g$-theorem. A physical realization of this setup occurs for instance in the Kondo problem, where $\log g$ measures the impurity entropy. The boundary entropy can also be obtained from the entanglement entropy on an interval $x_1 \in [0, R)$~\cite{Calabrese:2009qy},
\be
S(r) = \frac{c}{6}\,\log\frac{R}{\epsilon}+c_0 + \log g\,,
\ee
with $\epsilon$ a short distance cutoff and $c_0$ a bulk constant contribution that drops out from the differences $\log g_{UV} - \log g_{IR}$ we are interested in.

Let $\sigma$ be the density matrix of $BCFT_{UV}$ reduced to the interval $x_1 \in [0, R)$, and $\rho$ the corresponding quantity for the theory (\ref{eq:bRG}) with nontrivial RG flow. Introducing the modular Hamiltonian $\mc H = - \log \sigma$, the relative entropy can be written as
\be
S(\rho||\sigma)=\Tr (\rho \log \rho-\rho \log \sigma) = \Delta \langle \mc H \rangle -\Delta S\,,
\ee
where $\Delta \langle \mc H \rangle = \Tr \left((\rho-\sigma)\mc H \right)$, and $\Delta S = S(\rho) - S(\sigma)$. Since the relative entropy contains a piece that is the difference between entanglement entropies of the two theories, it is sensitive to the change in boundary entropy $\log (g(R)/g_{UV})$. However, in general the change in the modular Hamiltonian dominates in the relative entropy, with $\Delta \langle \mc H \rangle \propto R$.	

\begin{figure}[h!]
\begin{center}  
\includegraphics[width=0.5\textwidth]{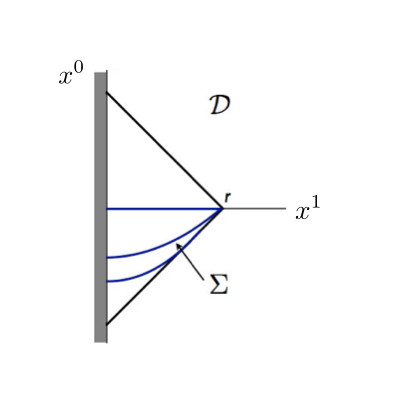}
\captionsetup{width=0.9\textwidth}
\caption{Choice of different Cauchy surfaces $\Sigma$ inside the interval $[0, r)$. As the Cauchy surface approaches the light-cone, the contribution $\Delta \langle \mc H \rangle \to 0$.
}
\label{fig:Cauchy}
\end{center}  
\end{figure}  

A direct connection between the relative entropy and the change in boundary entropy obtains by quantizing on a Cauchy surface $\Sigma$ that approaches the light-cone~\cite{Casini:2016fgb}. This is illustrated in Fig.~\ref{fig:Cauchy}. $\Delta S$ is independent of the choice of Cauchy surface, but $\Delta \langle \mc H \rangle$ depends on $\Sigma$ because the two density matrices evolve with different unitary operators. In the light-cone limit, $\Delta \langle \mc H \rangle \to 0$, and then
\be
S(\rho||\sigma)= \log \frac{g_{UV}}{g(R)}\,.
\ee
Positivity of the relative entropy then implies that $g(R)$ decreases monotonically under boundary RG flows, thus establishing the entropic $g$-theorem~\cite{Casini:2016fgb},
\be\label{eq:gthm}
\log g_{UV}-\log g_{IR} \ge 0\,.
\ee

In the present setup of boundary RG flows with Cauchy surface on the light-cone, the inequality (\ref{eq:RGbound1}) gives
\be\label{eq:bRGbound1}
S_\alpha(\rho||\sigma) \le  \log \frac{g_{UV}}{g(R)}
\ee
for a region $x_1 \in [0, R)$ with one endpoint at the boundary. In particular, in the IR limit $ mR \gg 1$, with $m$ the mass scale associated to the RG,
\be\label{eq:bRGbound2}
\lim_{m R \gg 1} S_\alpha(\rho||\sigma) \le  \log \frac{g_{UV}}{g_{IR}}\,.
\ee
The right hand side of this inequality is finite and depends only on the UV and IR fixed points, and not on the specific RG trajectory that connects them. On the other hand, we expect the left hand side to depend on the RG trajectory. 

Eq.~(\ref{eq:bRGbound2}) is our main result for boundary RG flows; it provides a bound on \textit{all possible} RG trajectories connecting two BCFTs. The upper bound $\log (g_{UV}/g_{IR})=-\Delta S$ depends only on intrinsic quantities of the fixed points. From a thermodynamic perspective, (\ref{eq:bRGbound2}) resembles the second law $\Delta S \ge \int \frac{dQ}{T}$. The thermal entropy is a function of state, while heat transfer depends on the process. In our present context, the ``function of state'' corresponds to the boundary entropy, which depends only on properties of the fixed point, while the QRRE, like heat transfer, is sensitive to the specific trajectory in coupling space. One interesting difference between the thermodynamic and quantum cases is that, while in the former case there exists nontrivial adiabatic processes with no exchange of heat, in the quantum setup ``adiabatic RG flows'' are not possible. In other words, $\log g_{UV} - \log g_{IR}$ is always strictly positive. This follows from the fact that this difference is a relative entropy, which vanishes only for $\rho = \sigma$. But if the two density matrices agree, then all correlators are the same, and there is no RG flow.\footnote{This is also a consequence of the formula found in~\cite{Friedan:2003yc}, which relates $\log (g_{UV}/g_{IR})$ to an integral of the two-point function of the stress-tensor trace. This two-point function vanishes if and only if the trace vanishes as an operator, in which case there is no RG flow.}

Finally, we note that as $m R \gg 1$, we expect the fidelity to approach the overlap of the corresponding vacuum wavefunctionals. The inequality (\ref{eq:bRGbound2}) for $\alpha=1/2$ then yields
\be
-2 \log\,|\langle \Psi_\rho | \Psi_\sigma \rangle| \le  \log \frac{g_{UV}}{g_{IR}}\,.
\ee
The wavefunctionals here are defined on the past (or future) null infinity. The finite right hand side implies a nonzero overlap $|\langle \Psi_\rho | \Psi_\sigma \rangle|\neq 0$. This is an interesting outcome, which provides a way of avoiding Anderson's orthogonality catastrophe in relativistic systems. Anderson's result~\cite{anderson1967infrared} states that, under mild assumptions, a many-body fermion ground-state wavefunction $|\Psi'\rangle$ in the presence of a local perturbation is orthogonal to the unperturbed ground state, $|\langle \Psi | \Psi' \rangle| =0$. In fact, if we work on a Cauchy surface at constant time, we expect the same result for $|\langle \Psi_\rho | \Psi_\sigma \rangle|$ in the more general boundary RG flows we are considering --we will see an example of this in Sec.~\ref{sec:Kondo}. However, the orthogonality is avoided by taking the light-cone limit. The finite overlap is guaranteed by the relative entropy becoming finite in this limit, and corresponds to both theories being less distinguishable on the light-cone. It would be interesting to understand other consequences of this result.

\subsection{RG flows in $d \ge 2$ dimensions}\label{subsec:RGd}

We will now consider RG flows in $d$ spacetime dimensions, where the fixed points --denoted by $CFT_{UV}$ and $CFT_{IR}$-- are Poincare invariant unitary CFTs. These flows can be produced by turning on relevant deformations in $CFT_{UV}$,
\be\label{eq:SRG}
S= S_{CFT_{UV}}+ \int d^dx\, g\, \mathcal O\;\;,\;\;\Delta_{\mc O} \le d\,,
\ee
with $\Delta_{\mc O}$ the scaling dimension of $\mc O$ at the UV fixed point.
The light-cone construction summarized in the previous section has been extended to this case in~\cite{Casini:2016udt}, and we will now examine the implications of (\ref{eq:RGbound1}).

We again introduce two reduced density matrices $\sigma$ and $\rho$, associated to the sphere $r \le R$; $\sigma$ corresponds to $CFT_{UV}$, while $\rho$ arises in (\ref{eq:SRG}). Each of them is obtained by starting from the vacuum state $|0 \rangle \langle 0|$ of the corresponding theory, and tracing over the degrees of freedom in the complement of sphere. The theories have the same operator content, but evolve with different hamiltonians. The QRREs $S_\alpha(\rho_V||\sigma_V)$ provide distance measures for the RG. We can evaluate them on different Cauchy surfaces inside the causal domain of dependence on $V$ (recall Fig.~\ref{fig:Cauchy}), and we will focus on surfaces that approach the light-cone.

Let us focus first on $d=2$ spacetime dimensions. Ref.~\cite{Casini:2016udt} showed that the modular Hamiltonian contribution to the relative entropy vanishes in the light-cone limit, as in the case of boundary RG flows. Denoting the characteristic mass scale of (\ref{eq:SRG}) by $m$, the relative entropy in the large distance limit $R \gg 1/m$ becomes~\cite{Casini:2016udt}
\be
S(\rho||\sigma) = - \Delta S \approx \frac{c_{UV}- c_{IR}}{3}\,\log(m R)\,,
\ee
where $c$ is the CFT central charge.
Positivity of the relative entropy then provides an alternative proof of the $c$-theorem, $c_{UV} \ge c_{IR}$.\footnote{The first proof of this theorem was given by Zamolodchikov, based on local QFT correlators~\cite{Zamolodchikov:1986gt}.} Combining this with (\ref{eq:RGbound1}), we find the following restriction on RG trajectories:
\be\label{eq:RGbound2}
\lim_{m R \gg 1}\,S_\alpha(\rho||\sigma) \le \frac{c_{UV}- c_{IR}}{3}\,\log(m R)\,.
\ee
As in (\ref{eq:bRGbound2}), we have here a function of the trajectory being bounded above by a quantity that is intrinsic to the fixed points. We conclude that the distances $S_\alpha(\rho||\sigma)$ can grow at most logarithmically at long distances, and with a coefficient that is smaller than $(c_{UV}-c_{IR})/3$.

In higher dimensions, the relative entropy in the light-cone limit is dominated by the area term $\mu_{d-2}$ in the entanglement entropy,
\be
S(R) = \mu_{d-2}\,R^{d-2}+ \ldots
\ee
(Note that in $d=2$, the central charge $c$ also appears as the leading area term.) For relevant deformations with dimension $\Delta_{\mc O} < (d+2)/2$, the modular Hamiltonian contribution to the relative entropy vanishes, and~\cite{Casini:2016udt}
\be
S(\rho||\sigma) \approx (\mu_{UV}-\mu_{IR}) R^{d-2}
\ee
in the limit $mR \gg 1$. Therefore, we arrive at the constraint
\be\label{eq:RGbound3}
\lim_{m R \gg 1}\,S_\alpha(\rho||\sigma) \le (\mu_{UV}- \mu_{IR})\,R^{d-2}\,.
\ee
Unlike the previous cases, $\mu_{UV} -\mu_{IR}$ is not a combination of intrinsic quantities. However, it is still an interesting object in QFT. It is given by the integral of the stress tensor correlator~\cite{Casini:2014yca, Casini:2015ffa}
\be\label{eq:sumrule}
\mu_{UV}-\mu_{IR} = - \frac{\pi}{d(d-1)(d-2)}\,\int d^dx\,x^2 \,\langle \Theta(x) \Theta(0) \rangle\,,
\ee
where $\Theta$ is the trace of the stress tensor. This quantity is finite when $\Delta_{\mc O} < (d+2)/2$, and depends on the RG trajectory.\footnote{For instance, we can deform a free boson or fermion by a mass term and then evaluate (\ref{eq:sumrule}). The result depends on the mass parameter, which is not intrinsic to the free fixed point.} It is also proportional to the renormalization of Newton's constant due to the field-theoretic degrees of freedom in flat space.

This ends our general discussion on constraints for RG flows. In the next section we will study the QRRE in a concrete Kondo model with a nontrivial flow.

\section{Application to the free Kondo model} \label{sec:Kondo}

Finally, we will study the distances $S_\alpha$ in the Kondo problem introduced in~\cite{Cardy:2016fqc}. This model is free but it supports a nontrivial boundary RG flow, providing an interesting setup where the $S_\alpha$ can be evaluated nonperturbatively in the relevant deformation.

\subsection{The free Kondo model}

The model consists of a free Dirac fermion $\psi$ living in half-space $x^1 \ge 0$. This `bulk' fermion is coupled quadratically to a fermionic degree of freedom $\chi$ that lives at the boundary $x^1=0$,
\be\label{eq:kondo1}
S= \int_{-\infty}^\infty dx^0 \int_0^\infty dx^1\, \left(-i \bar \psi \gamma^\mu\partial_\mu\psi+\frac{i}{2}\delta(x_1) \left[\bar\chi \gamma^0\partial_0 \chi+m^{1/2}(\bar \psi \chi-\bar\chi \psi)\right] \right)\,.
\ee
In the UV, we choose the boundary condition that relates the two chiralities,
\be\label{eq:bUV}
\psi_+(x^0, 0) =\psi_-(x^0, 0)\,.
\ee
In the IR, $E \ll m$, the mass term dominates over the impurity kinetic term, and extremizing over $\chi$ sets
\be\label{eq:bIR}
\psi_+(x^0, 0) =- \psi_-(x^0, 0)\,.
\ee
Hence we obtain a boundary RG flow between `$+$' and `$-$' boundary conditions for the Dirac fermion.

The lattice version of the theory contains a single fermion $\psi_j$ hopping in a one-dimensional lattice, with just one special site corresponding to the impurity,
\be\label{eq:lattice}
L= a\sum_{j=0}^\infty \left(i \psi_j^* \partial_0 \psi_j-\frac{i}{2a}(\psi_j^* \psi_{j+1}-\psi_{j+1}^* \psi_j)\right)+i \eta^* \partial_0 \eta-\frac{i}{2} m^{1/2}(\eta^* \psi_0+c.c.) \,.
\ee
Here $a$ is the lattice spacing, and $\eta$ is the impurity fermion. The spectrum of $\psi_i$ contains left and right moving low energy modes, as expected from the usual fermion doubling. Furthermore, constructing a Majorana fermion $\chi$ out of the lattice fermion $\eta$, the last term in (\ref{eq:lattice}) produces the quadratic coupling of (\ref{eq:kondo1}). In this way, the continuum limit of this lattice model reproduces (\ref{eq:kondo1}).

As a first step, let us evaluate the fidelity distance on a Cauchy surface at constant time. In the next section we will analyze the light-cone limit.


For this, we need to calculate the equal-time fermion two-point functions $C$ and $C'$. $C'$ is the correlator for arbitrary mass $m$, while $C$ arises for the particular case $m=0$. This calculation was described in detail in~\cite{Casini:2016fgb}, and for completeness here we summarize the main points.

The quadratic kernel $M_{ij}$ for the Hamiltonian in (\ref{eq:lattice}) can be diagonalized in terms of momentum modes $\psi_j(k)= e^{ikj}+ R_k (-1)^j e^{-i kj}$; the energies in units of $a=1$ read $E(k) = -\sin k$. The reflection coefficient $R(k)$ follows from the first two equations in the diagonalization of $M$, and becomes
\be
R(k) =-\frac{1-m-e^{-2 i k}}{1-m-e^{2ik}}\,.
\ee
Taking the continuum limit, $R(k) \to 1$ in the UV, and $R(k)  \to -1$ in the IR. This reproduces the boundary RG flow in (\ref{eq:bUV}) and (\ref{eq:bIR}). The equal time correlator is simply given by
\be\label{corr}
C_{ij}(m) =- \int \frac{dk}{2\pi}\,\Theta(-E(k))\,\psi_i^\dagger(k) \psi_j(k) = - \int_{0}^{\pi/2} \frac{dk}{2\pi} \psi_i^\dagger(k) \psi_j(k)\,.
\ee

Using then in \eqref{corr} the explicit expressions for the wavefunctions we can compute the fermionic correlators for the theories we want to compare and, with the help of \eqref{eq:SfinalF}, we finally evaluate numerically $S_{1/2}$.

\begin{figure}[h]
\center
\includegraphics[width=10cm]{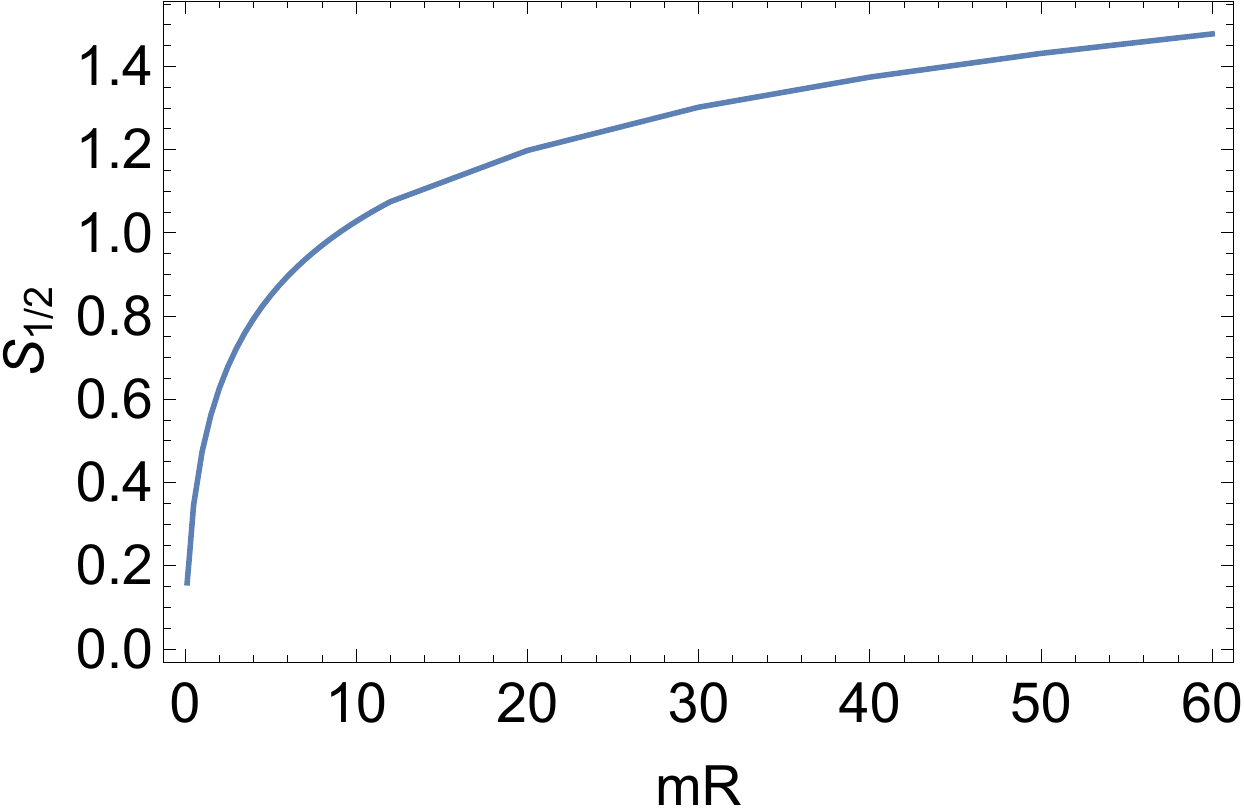}
\caption{\label{fig:fid}Quantum fidelity for the RG flow in the Kondo model, formulated on the $t=0$ line.}
\end{figure}

In Fig.~\ref{fig:fid} we show $S_{1/2}$ as a function of the interval size, for intervals with one end at the boundary. The increase of $S_{1/2}(mR)$ with $mR$ is rather slow, but our analysis shows that $S_{1/2}(mR) \to \infty$ as $mR\to \infty$. In other words, the fidelity $F(\rho, \sigma) \to 0$ in this limit. This is consistent with the bound (\ref{eq:RGbound1}), since $S_{1/2}(\rho||\sigma)$ grows more slowly than the relative entropy, which scales like $S_1(\rho||\sigma) \sim mR$ for large intervals~\cite{Casini:2016fgb}. Our numerical findings indicate then that the whole family of $S_\alpha$ diverges when $mR \to \infty$, since the QRREs are monotonically increasing in $\alpha$.

As stated by Uhlmann's theorem \eqref{fid2}, the fidelity can be defined as a maximization over purifications. For $mR \gg 1$, the system containing the impurity is becoming pure, and then the fidelity is given by the overlap of vacuum wavefunctions of the theory with and without mass perturbation. We can then understand the fact that $F(mR\to\infty)\to0$ as an expression of Anderson's orthogonality catastrophe~\cite{anderson1967infrared}, to the effect that the ground state in the presence of a local perturbation should become orthogonal to the original ground state in the thermodynamic limit.

\subsection{The Kondo model on the null line}

Given the previous results on a Cauchy surface at constant time, we will now analyze the measures $S_\alpha$ on regions lying on the null line with one end on the boundary. From the general discussion in Sec.~\ref{subsec:bRG}, this should give finite QRREs as $mR \gg 1$. The reason for this change in behavior as we modify the Cauchy surface is that we are comparing two density matrices $\sigma$ and $\rho$ that evolve with different Hamiltonians.

The fermion correlator on the null line takes the form~\cite{Casini:2016fgb}
\be
\label{eq:nullco}
C(mR) =\left( \begin{matrix} 
1/2 & a(s, mR) \\ a^*(s, mR) & \text{diag}(\lambda_s)
\end{matrix}\right)
\ee
where
\be
\lambda_s = \frac{1+\tanh(\pi s)}{2}
\ee
and
\be
a(s, mR) = \int_0^{mR} dz \frac{i(mR)^{1/2}}{(2\pi)^{3/2}}\,\frac{e^{z/2}\,\text{Ei}(-z/2)}{z^{1/2}(mR-z)^{1/2}}\, e^{-i s \log \frac{z}{mR -z}}\,.
\ee
Here $\text{Ei}$ is the exponential integral function
\be
\text{Ei}(x)=-\int_{-x}^\infty dt \frac{e^{-t}}{t}\,.
\ee
In these expressions, $-\infty< s < \infty$.\footnote{To be precise, here $s$ does not contain $0$. This is taken into account by the first element of the matrix, since $\lambda_0=1/2$. Anyway, $a(s)$ will always appear inside integrals, and subtracting the contribution from $s=0$ does not change the results.}

In order to compute the QRRE, we evaluate the resolvents defined in (\ref{eq:SfinalF2}). This requires calculating a few inverses and powers of correlators. The density matrix $\sigma$ corresponds to a correlator (\ref{eq:nullco}) with $m=0$ (this is the UV fixed point), while for $\rho$ we need to take an arbitrary $m$. This computation is quite lengthy but straightforward, and we detail the steps in Appendix \ref{apb}. The resulting expression for the QRRE is
\bea
\label{eq:final}
S_\alpha( \rho|| \sigma)&=&-\frac\alpha{1-\alpha}\log2+\frac{1}{ \pi }\frac{1}{1-\alpha} \int_0^\infty d\beta\, \frac{ \alpha \beta^{\alpha-1} \sin(\pi \alpha)   }{1+2\beta^\alpha \cos(\pi\alpha)+\beta^{2\alpha}}  \Bigg\lbrace- \log\left(\beta+1\right)\;\;\; \nonumber\\
&+&\log\left[ (\beta-1)\left(\frac{1}{2}+ \int ds |a(s)|^2 \left( \frac{\cosh^2(\pi s)}{1-e^{-2\pi s/\alpha}\beta}-\frac{2}{1+\tanh \pi s}\right)\,\right)+1 \right]\Bigg\rbrace \;\;
\eea

\begin{figure}[h]
\center
\includegraphics[width=10cm]{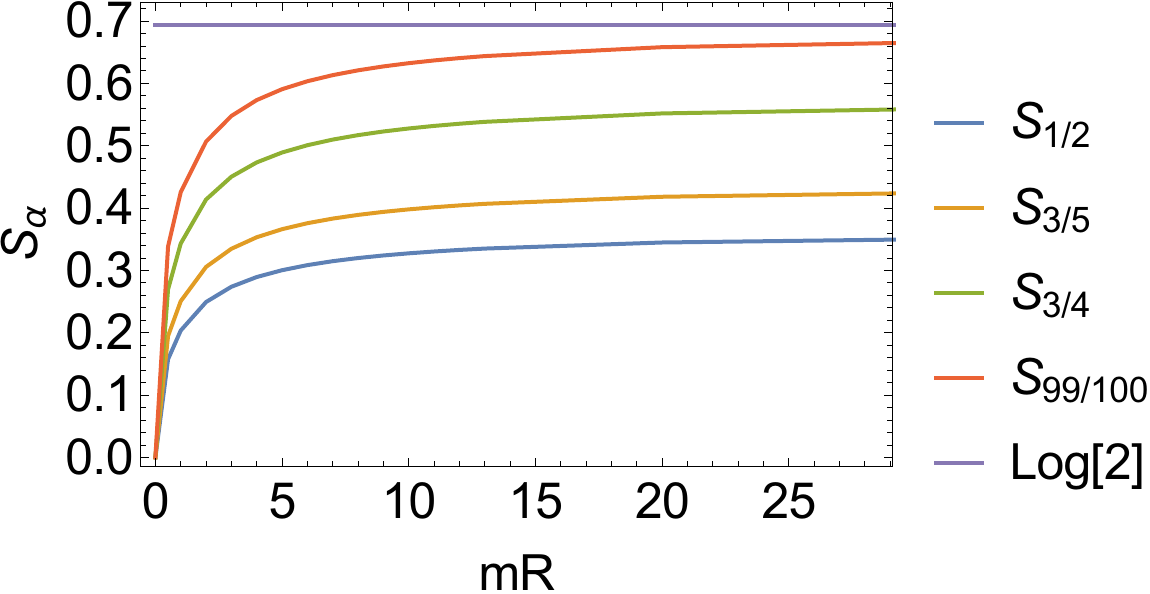}
\caption{\label{fig:rre} QRREs $S_\alpha(\rho||\sigma)$ for the RG flow in the Kondo model, formulated on the light-cone. We show different values of $\alpha$, and the limiting $\log 2$ result for $mR \gg 1$ and $\alpha \to 1$.}
\end{figure}

In Fig.~\ref{fig:rre} we show the result of evaluating (\ref{eq:final}) numerically  for several values of $1/2\leq\alpha<1$. As predicted from the general properties of QRRE, the curves are monotonically increasing  as we increase the region size $mR$. Also, we observe the monotonicity in $\alpha$, with $S_\alpha<S_{\alpha'}$ for $\alpha<\alpha'$ and for all values $mR>0$. In the limit $mR\rightarrow 0$, all the $S_\alpha$ collapse to zero, consistent with $S_\alpha(\rho||\rho)=0$. These curves provide measures of distances between the states $\sigma$ and $\rho$ along the RG, with $m R \to 0$ corresponding to the UV (high energies), while $m R \to \infty$ approaches the IR limit.

\begin{figure}[h]
\center
\includegraphics[width=10cm]{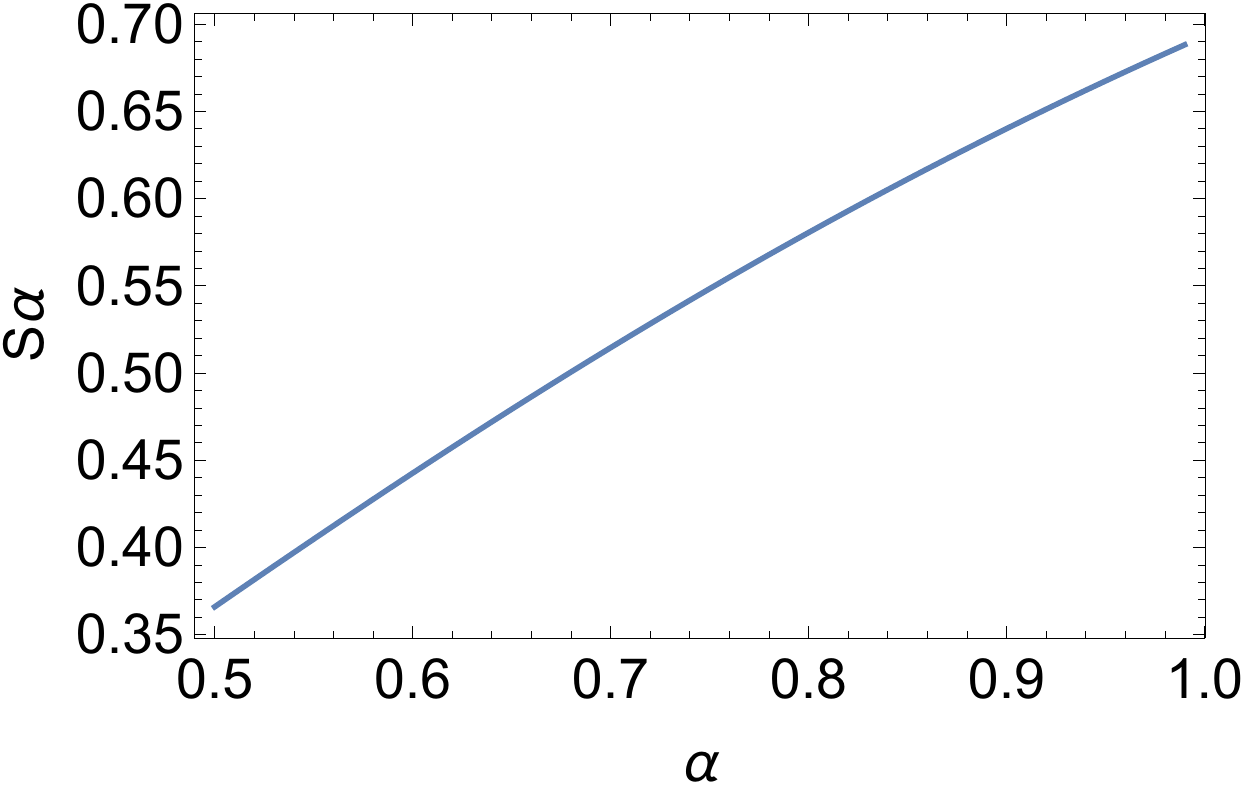}
\caption{\label{fig:mr} $S_\alpha$ in the  $m R \gg 1 $ limit, for different values of $\alpha$. }
\end{figure}

Finally, let us evaluate the limit $mR \gg1$.
In this limit, the integral for $a(s)$ approximates to
\be
a(s) \approx e^{i s \log mR}\,\frac{i}{(2\pi)^{3/2}}\,\int_0^\infty dz\, \frac{1}{z^{1/2+is}}\,e^{z/2}\,\text{Ei}(-z/2)\,.
\ee
The prefactor $e^{i s \log mR}$ drops out from all the expressions, since only $|a|^2$ enters. The integral can now be performed analytically, giving
\be
|a(s)|^2 = \frac{1}{4}\,\text{sech}^3(\pi s)\,.
\ee
Fig.~\ref{fig:mr} shows the results for $S_\alpha$ when $m R\gg 1$. In particular, we find that these distances asymptote to different values which depend on $\alpha$. For $\alpha=1$ we recover the result $\log (g_{UV}/g_{IR})=\log 2$ for the change in the impurity entropy~\cite{Casini:2016fgb}. On the other hand, for $\alpha<1$ we expect $S_\alpha$ to depend on the RG flow in between the fixed points. Finally, for $\alpha=1/2$, $e^{-S_{1/2}}$ measures the overlap between the wavefunctionals with and without perturbation. We obtain a finite result on the null Cauchy surface, providing a way to avoid Anderson's orthogonality catastrophe for this system.

\section{Conclusions and future directions} 
\label{sec:concl}

In this work we studied the quantum Renyi relative entropies for RG flows in quantum field theory. These measures provide interesting distances that characterize new aspects of the RG. We presented explicit expressions for the QRREs in free field theories, focusing in particular on fermionic systems. In this case, we studied a nontrivial RG flow in a Kondo model, and evaluated numerically the $S_\alpha$. The results on the light-cone limit are finite, show the irreversibility of the RG, and are expected to depend on the trajectory in coupling space.

Using QIT properties of the $S_\alpha$, we obtained constraints that consistent RG trajectories need to obey. These are strongest for boundary RG flows and flows in two spacetime dimensions -- we showed that the measures $S_\alpha$ are bounded above by differences of quantities that are intrinsic to the fixed points (impurity entropy or central charge). In higher dimensions, the bound is given by the area term in the entanglement entropy, which also measures the renormalization of Newton's constant. For boundary RG flows, and flows in two spacetime dimensions, these constraints closely resemble the second law of thermodynamics. This hints towards the thermodynamic nature of the RG, and it would be interesting to develop this analogy further.

Our analysis suggests several directions to explore. At the level of concrete examples, one can generalize the Kondo model to include more impurities with various relevant parameters. This would provide dimensionless couplings that parametrize different RG trajectories, and we expect a nontrivial dependence of the QRREs on such parameters. More generally, it would be useful to find a more direct connection between a given RG trajectory and the $S_\alpha$, perhaps in the form of a sum rule as in (\ref{eq:sumrule}). It would be interesting to study RG flows in other number of dimensions and with defects of different codimension. It would also be important to develop tools to evaluate these distances for more general CFTs and their RG flows, for instance using conformal perturbation theory. In this direction, the euclidean approach could provide a new handle on the problem. Holographic duals of these measures should also give insights into their physical properties.

\section*{Acknowledgments}
We thank M. Wilde for bringing various references to our attention after we submitted the first version of the preprint to the arXiv.
This work was partially supported by CONICET (PIP grant 11220150100299), CNEA, and Universidad Nacional de Cuyo, Argentina. H.C. acknowledges an ``It From Qubit'' grant of the Simons Foundation. G.T. is also supported by ANPCYT PICT grant 2015-1224.

\appendix

\section{$S_\alpha$ for Gaussian states}
\label{app:gaussian}

In this Appendix we present explicit calculations of $S_\alpha(\rho||\sigma)$ for Gaussian states. For the purpose of this paper, they arise as the vacuum-reduced density matrices of free QFTs, but they also occur in more general setups, where the Hamiltonians are not necessarily quadratic.

\subsection{Fermions}
\label{app:f}

In this section we focus on fermionic Gaussian states, and derive \eqref{eq:SfinalF}. A similar procedure was used for the fidelity in~\cite{banchi2014quantum}.

Let us consider a system of fermionic modes $\psi_i, \psi^\dagger_j$ described by a set Majorana operators $w_I=(\psi_j+\psi^\dagger_j, i(\psi_j-\psi^\dagger_j))$. In terms of these variables, the two point correlation function is $\mathcal{C}_{IJ}= \frac{1}{2} \langle [w_I, w_J]\rangle$. The complex matrix $\mathcal{C}$ is imaginary and anti-symmetric. 
Now, let us consider a gaussian fermionic state written in the form
\be\label{eq:rhoG}
\rho = \frac{1}{Z}\exp \left( -\frac{i}{4} \sum_{IJ} G_{IJ} w_I w_J \right)\,,
\ee
with $G$ real and antisymmetric. It is possible then, to cast G in the canonical form by an orthogonal matrix $O$
\begin{equation}
\label{linealalg}
G = O^T \bigoplus_{k = 1} \left( \begin{matrix}
						0 & g_k \\
						-g_k & 0
\end{matrix}
\right) O\,,
\end{equation}
with $\pm i g_k$ the eigenvalues of $G$. Now, let $r_I = \sum_{K} Q_{IK} w_K$ be the new Majorana operators. In this new basis, we find the following expression for the state $\rho$
\begin{equation}
\label{state2}
\rho = \frac{1}{Z} \prod_{k=1} \left( \cosh \left(\frac{g_k}{2}\right) - i \sinh\left(\frac{g_k}{2}\right) r_{2k-1}r_{2k} \right).
\end{equation}

The value of the normalization constant Z is fixed by requiring $\tr \rho = 1$,
\begin{equation}
\label{trace}
\tr \rho = 1 \Rightarrow Z = \sqrt{\det \left[ 2 \cosh\left(i\frac{G}{2}\right) \right]},
\end{equation}
where we used the fact that the eigenvalues of $i G$ are $\pm g_k$. $G$ and the correlation matrix are related by
\begin{equation}
\label{CorrFunc}
\mathcal{C} = -\frac{4 i}{Z} \frac{\partial Z}{\partial G} = \tanh \left( \frac{i G}{2} \right) \,.
\end{equation}

Let us now evaluate the $S_\alpha(\rho||\sigma)$ for states of the form (\ref{eq:rhoG}). Recalling (\ref{eq:prodf}), it follows that
\begin{equation}
\label{powstate}
\left( \sigma^{\frac{1-\alpha}{2 \alpha}} \rho \sigma^{\frac{1-\alpha}{2 \alpha}} \right)^\alpha  \propto \exp \left(  \frac{\alpha}{4} \sum_{IJ}  \log \left( \e^{-i \frac{1-\alpha}{2 \alpha} G }\e^{-i G' }\e^{-i \frac{1-\alpha}{2 \alpha} G } \right)_{IJ} w_I w_J \right)\,.
\end{equation}
Finally, using \eqref{powstate}, \eqref{eq:prodf} and \eqref{trace} we find for the QRRE
\begin{equation}
\label{Sversion1F}
S_\alpha (\rho || \sigma) = -\frac{1}{1-\alpha}\frac{\det \big[ \cosh \big( \frac{\alpha}{2} \log \big( \e^{-i \frac{1-\alpha}{2 \alpha} G }\e^{-i G' }\e^{-i \frac{1-\alpha}{2 \alpha} G } \big) \big) \big]^{1/2}}{\Big(\sqrt{\det \big[\cosh(i\frac{G}{2}) \big]} \Big)^{1-\alpha} \Big(\sqrt{\det \big[\cosh(i\frac{G'}{2}) \big]} \Big)^{\alpha}}\,.
\end{equation}

In order to express the QRRE as a function of the fermionic correlators we define the following convenient parametrization
\begin{equation}
\label{param}
T = e^{i G}, \hspace{1cm} \mathcal{C}=\frac{T-1}{T+1},\hspace{1cm} T^T = T^{-1}, \hspace{1cm} T^\dagger = T\,,
\end{equation}
in terms of which
\begin{equation}
\label{SversionT}
S_\alpha (\rho || \sigma) = -\frac{1}{1-\alpha}\log \frac{\det \big[1 + \big( T^{ \frac{1-\alpha}{2 \alpha} }T'T^{ \frac{1-\alpha}{2 \alpha} } \big)^\alpha]^{1/2}}{\Big(\sqrt{\det [1+T]} \Big)^{1-\alpha} \Big(\sqrt{\det [1+T']} \Big)^{\alpha}}.
\end{equation}

Lastly, we take into account that we are interested in models with charge conjugation symmetry, which fixes $\text{Re}( C_{ij} )=\frac{1}{2}\delta_{ij}$. The matrix $\mathcal{C}$ becomes 
\begin{equation}\label{Cmatrix}
\mathcal{C} =\left( \begin{matrix}
	2 \,\text{Im} (C) & 0 \\
	0 & 2 \,\text{Im} (C)
\end{matrix} 
\right).
\end{equation} 
Using \eqref{Cmatrix} and \eqref{SversionT} we arrive at our final result
\bea
S_\alpha(\rho||\sigma) & =& -\Tr \log(1-C) - \frac{\alpha}{1-\alpha}\Tr \log(1-C') \\
& -&\frac{1}{1-\alpha}\Tr \log \left[ 1 +\left( \Big(\frac{C}{1-C} \Big)^{\frac{1-\alpha}{2 \alpha}} \frac{C'}{1-C} \Big(\frac{C}{1-C} \Big)^{\frac{1-\alpha}{2 \alpha}}\right)^{\alpha} \right] \nonumber\,.
\eea

\subsection{Bosons}
\label{app:b}

For bosons, we shall discuss Gaussian states of the form
\be
\rho_V \propto \exp\left(-\sum_{i,j \in V} \big( M_{ij} \phi_i \phi_j + N_{ij}\pi_i \pi_j \big) \right)\,,
\ee
with vanishing $\phi \pi $ terms. This appears naturally in systems with time-reversal invariance. Free bosonic QFTs of the form discussed in Sec.~\ref{subsec:bosons} are a special case; see (\ref{eq:rhoboson-general}).

Performing a Bogoliubov transformation as in~\cite{Casini:2009sr}, obtains
\be\label{bosonstate}
\rho = \frac{1}{Z} e^{-\frac{1}{2}Q^T G Q}\,,
\ee
where $Q_I = (\phi_j,\pi_j)$ and 
\be\label{eq:WillG}
G=S\, \text{diag}(\epsilon, \epsilon) S^T\,,
\ee 
with $S$ a symplectic matrix, namely $S^T \Omega S = \Omega$.\footnote{This last expression is called Williamson' normal form for the matrix $G$. This diagonalization applies to any square, positive-definite real matrix.} It is also useful to introduce the covariance matrix
\be\label{MeanCovariance}
V_{IJ} = \frac{1}{2} \langle \lbrace Q_I, Q_J \rbrace \rangle_\rho=\left( \begin{matrix}
	X & 0 \\
	0 & P
\end{matrix}
\right)\,.
\ee
The Bogoliubov transformation that diagonalizes $G$ also diagonalizes $V$,  
\be\label{eq:WillV}
V=S' \,\text{diag}(\nu,\nu)\,S'^T\,,
\ee 
where $\lbrace \nu_k \rbrace$ are the eigenvalues of the correlation matrix $C=\sqrt{XP}$, and the matrices in (\ref{eq:WillG}) and (\ref{eq:WillV}) are related by $S'= \Omega S$. Furthermore, the corresponding eigenvalues obey
\be\label{sympaction}
\nu_k (\epsilon_k) = \frac{1}{2} \coth\left(\frac{\epsilon_k}{2}\right)\,.
\ee 
This can also be seen by looking at a single bosonic mode. Since $G$ and $\Omega V \Omega$ are diagonalized by the same symplectic matrix $S$ (since $S'= \Omega S$), Eq.~(\ref{sympaction}) can be written as a matrix identity
\be\label{relationsGC}
V=\frac{1}{2}\coth\left(\frac{i \Omega G}{2}\right)i \Omega \,.
\ee
See also~\cite{banchi2015quantum} for a derivation in terms of symplectic actions.

Now let's concentrate on the normalization factor in \eqref{bosonstate},
\be\label{eq:Z}
Z= \Tr \,e^{-\frac{1}{2}Q^T G Q}\,.
\ee
When $G$ is diagonal (i.e. $V$ is diagonal) then 
\begin{equation}\label{partition}
Z = \prod_i \left( \frac{1}{e^{\frac{\epsilon_i(\nu)}{2}}-e^{-\frac{\epsilon_i(\nu)}{2}}} \right) = \prod_i \sqrt{\nu_i ^2 -1/4 } = \det\left[V_D+\frac{i\Omega}{2}\right]^{1/2},
\end{equation}
where $V_D =\text{ diag}(\nu,\nu)$. This is invariant under the symplectic transformation (\ref{eq:WillV}), and hence in the general nondiagonal case
\begin{equation}
Z = \det\left[V+\frac{i\Omega}{2}\right]^{1/2}.
\end{equation}
It is also possible to write $Z$ in the following form
\begin{equation}\label{partfinal}
Z = \frac{1}{\sqrt{\det[2\sinh(\frac{i \Omega G}{2})i\Omega]}}.
\end{equation}

We are now ready to compute the QRRE. In terms of the $Q$ variables, and taking into account normalization factors,
\be\label{eq:Sa1}
\Tr \left(\sigma^{\frac{1-\alpha}{2\alpha}} \rho \sigma^{\frac{1-\alpha}{2\alpha}} \right)^\alpha = \frac{\Tr \left(e^{-\frac{1}{2} \frac{1-\alpha}{2\alpha} Q^T G Q}e^{-\frac{1}{2} Q^T G' Q}e^{-\frac{1}{2} \frac{1-\alpha}{2\alpha} Q^T G Q} \right)^\alpha}{\det[2\sinh(\frac{i \Omega G}{2})i\Omega]^{-\frac{1-\alpha}{2}} \det[2\sinh(\frac{i \Omega G'}{2})i\Omega]^{-\frac{\alpha}{2}}}\,.
\ee
The product of Gaussian states in (\ref{eq:Sa1}) is performed using (\ref{eq:prodgauss}), yielding an expression of the form $\Tr\, e^{-\frac{1}{2}Q^T G'' Q}$ with
\be
i\Omega G'' = \alpha \log e^{i \frac{1-\alpha}{2\alpha} \Omega G}e^{i \Omega G'}e^{i \frac{1-\alpha}{2\alpha}\Omega G}\,.
\ee
This last trace is again a partition function of the form (\ref{eq:Z}), and can be evaluated in terms of (\ref{partfinal}) and the corresponding matrix $G''$. Putting these results together we arrive at
\begin{equation}\label{Fidel}
S_\alpha (\rho || \sigma ) = -\frac{1}{1-\alpha} \log \frac{\det[\sinh(\frac{i \Omega G}{2})i\Omega]^{\frac{1-\alpha}{2}} \det[\sinh(\frac{i \Omega G'}{2})i\Omega]^{\frac{\alpha}{2}}}{\det \Big[\sinh \Big(\frac{\alpha}{2} \log e^{i \frac{\Omega G}{2}}e^{i G'}e^{i \frac{\Omega G}{2}} \Big)i \Omega \Big]^{1/2}}\,.
\end{equation}

It is convenient to introduce $T = e^{i \Omega G}$, and rewrite
\begin{equation}\label{FidelT}
S_\alpha(\rho || \sigma) = -\frac{1}{1-\alpha} \log \frac{\det \big[ T-1 \big]^{\frac{1-\alpha}{2}} \det \big[ T'-1 \big]^{\frac{\alpha}{2}}}{\det \Big[(T^{\frac{1-\alpha}{2\alpha}}T'T^{\frac{1-\alpha}{2\alpha}})^{\alpha}-1 \Big]^{1/2}}\,.
\end{equation}
Using the relations \eqref{relationsGC} and the definition for $T$ we find
\begin{equation}\label{detT}
\det \big[ T-1 \big] = \frac{1}{\det \left[V i \Omega -\frac{1}{2}\right]} = \frac{1}{\det \left[\frac{1}{4} -C^2\right]}\,.
\end{equation}
It only remains to simplify the term with non trivial powers of T, $(T^{\frac{1-\alpha}{2\alpha}}T'T^{\frac{1-\alpha}{2\alpha}})^{\alpha}$. For this purpose let us study the structure of the $T$ matrices. Using again \eqref{relationsGC} and $T = e^{i \Omega G}$ we find that
\begin{equation}\label{Tmatrix}
T = \frac{V + \frac{i \Omega}{2}}{V - \frac{i \Omega}{2}}=\left( \begin{matrix}
X & i/2 \\
-i/2 & P
\end{matrix}
\right)\left( \begin{matrix}
X & -i/2 \\
i/2 & P
\end{matrix}
\right)^{-1}.
\end{equation}
Computing the inverse of the matrix and taking the matrix product obtains
\begin{equation}\label{FinalT}
\begin{split}
T &= \left( \begin{matrix}
X & i/2 \\
-i/2 & P
\end{matrix}
\right)\left( \begin{matrix}
P \frac{1}{C^2-\frac{1}{4}} & \frac{i}{2} P \frac{1}{C^2-\frac{1}{4}}P^{-1} \\
-\frac{i}{2} \frac{1}{C^2-\frac{1}{4}} & P^{-1} + \frac{1}{4} \frac{1}{C^2-\frac{1}{4}}P^{-1}
\end{matrix}
\right) \\
& = \left( \begin{matrix}
\frac{C^2+\frac{1}{4}}{C^2-\frac{1}{4}} & i \frac{C^2}{C^2-\frac{1}{4}}P^{-1} \\
-i P \frac{1}{C^2-\frac{1}{4}} & P \frac{C^2+\frac{1}{4}}{C^2-\frac{1}{4}}P^{-1}
\end{matrix}
\right)\,.
\end{split}
\end{equation}

Substituting \eqref{detT} into \eqref{FidelT} we arrive to the desired result for $S_\alpha$,
\be
\begin{split}
S_\alpha (\rho || \sigma) & =\frac{1}{2} \Tr \log \left( \frac{1}{4}-C^2 \right) + \frac{1}{2}\frac{\alpha}{1-\alpha} \Tr \log \left( \frac{1}{4}-C'^2 \right) \\
& + \frac{1/2}{1-\alpha} \Tr \log \left( \left( T^{ \frac{1-\alpha}{2 \alpha} }T'T^{ \frac{1-\alpha}{2 \alpha} } \right)^\alpha - 1 \right),
\end{split}
\ee
where $T$ is given by \eqref{FinalT}. Note that the result depends not only on $C$, but also on $P$ (or $X$).

\section{QRRE in the free Kondo model}
\label{apb}

In this Appendix we will give some details on the computation of the resolvents and the $S_\alpha$ for our free Kondo model on the null line.

Let us write the correlator \eqref{eq:nullco} in tensor notation as
\be
C'_{mn}= \lambda_m\,\delta_{mn}+ u_m a_n^*+ a_m u_n\,.
\ee
Here $u_m = \delta_{m0}$, and the indices $m, n = -\infty, \ldots, \infty$ (including $0$). We note the properties
\be
u^2=1\;,\; u \cdot a=0\;,\;u \cdot \lambda= 1/2\,.
\ee
The inverse matrix reads
\be\label{eq:Cinv}
(C'^{-1})_{mn}= \left(\gamma^{-1}-\frac{1}{\lambda \cdot u} \right) u_m u_n-\gamma^{-1}\left(u_m \frac{a_n^*}{\lambda_n} +u_n \frac{a_m}{\lambda_m} \right)+\lambda_m^{-1}\delta_{mn}+\gamma^{-1}\,\frac{a_m}{\lambda_m} \frac{a_n^*}{\lambda_n}\,,
\ee
where we have defined
\be
\gamma \equiv \lambda \cdot u - \sum_n \,\frac{|a_n|^2}{\lambda_n}\,.
\ee
This inverse can be obtained by proposing a linear combination of rank two tensors and then fixing the coefficients so that $C^{-1} C = \mathbf 1$. 
With this result, we compute the resolvent by inverting $C'-\beta$, finding
\be
R(C', \beta) =\frac{1}{(\lambda-\beta)\cdot u- \sum_n \,\frac{|a_n|^2}{\lambda_n-\beta}}\left( 1+\sum_m \,\frac{|a_m|^2}{(\lambda_m-\beta)^2}\right)-\frac{1}{(\lambda-\beta)\cdot u}+\sum_m \left( \frac{1}{\lambda_m-\beta}+\frac{1}{\beta}\right)\,.
\ee
All the sums are shorthand for integrals. Note that $(C)_{mn}=\lambda_m \delta_{mn}$.

Now we come to the calculation of $M$ 
\be
M= \left(\frac{C}{1-C}\right)^{\frac{1-\alpha}{2\alpha}}\,\frac{1}{1-C'}\,\left(\frac{C}{1-C}\right)^{\frac{1-\alpha}{2\alpha}}-\left(\frac{C}{1-C}\right)^{\frac{1-\alpha}{\alpha}}\,
\ee
and its resolvent. Computing $(1-C')^{-1}$ using (\ref{eq:Cinv}), we have
\be
(M-\beta)_{pq}= \left(\gamma^{-1}-2\right) u_p u_q+\gamma^{-1}(u_p \tilde a_q^*+u_q \tilde a_p)+ \left(\left(\frac{\lambda_p}{1-\lambda_p} \right)^{1/\alpha}-\beta \right) \delta_{pq}+\gamma^{-1}\,\t a_p \t a^*_q\,.
\ee
Here we defined
\be
\t a_p \equiv \left(\frac{\lambda_p}{1-\lambda_p} \right)^{\frac{1-\alpha}{2\alpha}}\,\frac{a_p}{1-\lambda_p}\,.
\ee
We also used the fact that $\gamma$ is invariant under $\lambda_n \to 1-\lambda_n$, using the explicit expressions for $\lambda_n$ and $a_n$. There were also factors of the form $\lambda \cdot u /(1-\lambda \cdot u)$ that simplify to $1$ recalling that $\lambda \cdot u=1/2$.

We proceed as before, proposing an inverse
\be
(M-\beta)^{-1}_{\quad qn}=\left(\left(\frac{\lambda_p}{1-\lambda_p} \right)^{1/\alpha}-\beta \right)^{-1}\delta_{qn}+ \bar \alpha u_q u_n+\bar \beta(u_q\,f_n \t a_n^*+ u_n f_q \t a_q)+ \eta\,f_q \t a_q \,f_n \t a^*_n
\ee
and fixing the coefficients. The result is
\bea
f_p&=& \left(\left(\frac{\lambda_p}{1-\lambda_p} \right)^{1/\alpha}-\beta \right)^{-1} \nonumber\\
\bar \alpha &=& \frac{1}{\beta-1}-\frac{\gamma+\sum_n f_n |\t a_n|^2}{(\beta+1)(\gamma+\sum_n f_n |\t a_n|^2)-1} \nonumber\\
\bar \beta&=& \frac{1}{(\beta+1)(\gamma+\sum_n f_n |\t a_n|^2)-1}\nonumber\\
\eta&=&-\frac{\beta+1}{(\beta+1)(\gamma+\sum_n f_n |\t a_n|^2)-1}\,.
\eea

The trace of the inverse can now be easily evaluated to yield
\be
\Tr\,\frac{1}{M-\beta}= \sum_n\left(\left(\frac{\lambda_n}{1-\lambda_n} \right)^{1/\alpha}-\beta \right)^{-1}+\bar \alpha +\eta\,\sum_n f_n^2 |\t a_n|^2\,,
\ee
using $u^2=1$ and $u \cdot \t a=0$. Putting everything together,
\be
R(M, \beta) = \int ds \left(\frac{1}{e^{2\pi s/\alpha}-\beta}+\frac{1}{\beta} \right)-\frac{1}{1-\beta}-\frac{\gamma+\sum_n f_n |\t a_n|^2+(\beta+1)\sum_n f_n^2 |\t a_n|^2}{(\beta+1)(\gamma+\sum_n f_n |\t a_n|^2)-1}
\ee
with
\bea
\gamma&=&\frac{1}{2}-\int ds\,\frac{|a(s)|^2}{\frac{1+\tanh \pi s}{2}} \nonumber\\
\sum_n f_n |\t a_n|^2&=&4\int ds\,\frac{\cosh^2(\pi s)}{1-e^{-2\pi s/\alpha}\beta}\,|a(s)|^2\nonumber\\
\sum_n f_n^2 |\t a_n|^2&=&4\int ds\,\frac{\cosh^2(\pi s)}{(e^{\pi s/\alpha}-e^{-\pi s/\alpha}\beta)^2}\,|a(s)|^2\,.
\eea

It is clear that for $mR=0$, $S_\alpha( C|| C)=0$. We can then subtract the $mR=0$ answer to the finite $mR$ expression, term by term, and this will make the $s$ integrals explicitly finite. The result is
\bea\label{eq:Salpha2}
S_\alpha(\rho|| \sigma)&=&-\frac{\alpha}{1-\alpha} \int_1^\infty d\beta \,\left( \frac{1}{1/2-\beta-\int \frac{|a_s|^2}{\lambda_s-\beta}}\left(1+\int \frac{|a_s|^2}{(\lambda_s-\beta)^2}\right)-\frac{1}{1/2-\beta} \right)\\
&-&\frac{1}{1-\alpha} \int_0^\infty d\beta\, \frac{\text{Im}\,\log(1+\beta^\alpha e^{i \pi \alpha}) }{\pi} \left( \frac{1}{1+\beta}+\frac{\gamma+\sum_n f_n |\t a_n|^2+(-\beta+1)\sum_n f_n^2 |\t a_n|^2}{(-\beta+1)(\gamma+\sum_n f_n |\t a_n|^2)-1}\right) \nonumber\,.
\eea
The sums (integrals) of $\t a$ in the last line are to be evaluated at $-\beta$.

We will now use 
\be
\partial_\beta \sum_n f_n |\t a_n|^2 = - \sum_n f_n^2 |\t a_n|^2\,,
\label{eq:use}
\ee
to integrate by parts in (\ref{eq:Salpha2}). The result is
\bea
S_\alpha(\rho|| \sigma)&=&\frac{1}{ \pi }\frac{-1}{1-\alpha} \int_0^\infty d\beta\, \frac{ \alpha \beta^{\alpha-1} \sin(\pi \alpha)   }{1+2\beta^\alpha \cos(\pi\alpha)+\beta^{2\alpha}}  \Bigg\lbrace \log\left(1+\beta\right) \nonumber\\
&-&\log\left( (-1+\beta)\left(\gamma+ \sum_n f_n |\t a_n|^2\right)+1 \right)\Bigg \rbrace -\frac\alpha{1-\alpha}\log2\,.
\eea
After changing the formal sums for their continuum limit integrals, this is the expression (\ref{eq:final}) appearing in the main text.

\bibliography{EE}{}
\bibliographystyle{utphys}

\end{document}